\documentclass[conference]{IEEEtran}
\IEEEoverridecommandlockouts
\usepackage[most]{tcolorbox}
\usepackage{amsmath}
\usepackage{amsthm}
\usepackage{url}
\usepackage{tikz}
\usepackage{multirow}

\usepackage{colortbl}

\usepackage{graphicx}
\usepackage{xcolor}
\usepackage{color}

\usepackage{varwidth}
\usepackage{rotating}

\usepackage[none]{hyphenat} 

\usepackage{nameref}
\usepackage{zref-xr,zref-user}
\zxrsetup{toltxlabel=true, tozreflabel=false}
\usepackage{xcite}

\usepackage{tikz-3dplot}

\usepackage{subfigure}

\usetikzlibrary{decorations.pathreplacing,decorations.markings}

\tikzset{arrow data/.style 2 args={%
      decoration={%
         markings,
         mark=at position #1 with \arrow{#2}},
         postaction=decorate}
      }%

\usetikzlibrary{shapes,backgrounds,calc}

\tikzstyle{dummy} = [rectangle, text width=0.1em, draw=white, white,
                      minimum width=0.1em, minimum height=3em, opacity=0.0]

\makeatletter
\newcommand*{\Strut}[1][0.1em]{\vrule\@width\z@\@height#1\@depth\z@\relax}
\makeatother

\begin{document}

\title{Will the Proliferation of 5G Base Stations Increase the Radio-Frequency ``Pollution''?}
\author{Luca Chiaraviglio,$^{1,2}$ Giuseppe Bianchi,$^{1,2}$ Nicola Blefari-Melazzi,$^{1,2}$ Marco Fiore,$^{3}$\\
(1) Department of Electronic Engineering, \\University of Rome Tor Vergata, Rome, Italy, email \{luca.chiaraviglio,giuseppe.bianchi,blefari\}@uniroma2.it\\
(2) Consorzio Nazionale Interuniversitario per le Telecomunicazioni, Italy\\
(3) Institute of Electronics, Computer and Telecommunication Engineering, \\National Research Council of Italy, Turin, Italy, email marco.fiore@ieiit.cnr.it 
}

\maketitle

\IEEEpeerreviewmaketitle

\begin{abstract}
A common concern among the population is that installing new 5G Base Stations (BSs) over a given geographic region may result in an uncontrollable increase of Radio-Frequency ``Pollution'' (RFP). To face this dispute in a way that can be understood by the layman, we develop a very simple model, which evaluates the RFP at selected distances between the user and the 5G BS locations. We then obtain closed-form expressions to quantify the RFP increase/decrease when comparing a pair of alternative 5G deployments. Results show that a dense 5G deployment is beneficial to the users living in proximity to the 5G BSs, with an abrupt decrease of RFP (up to three orders of magnitude) compared to a sparse deployment. We also analyze scenarios where the user equipment minimum detectable signal threshold is increased, showing that in such cases a (slight) increase of RFP may be experienced.
\end{abstract}

\begin{IEEEkeywords}
5G Cellular Networks; Radio-Frequency Pollution; Base Station Deployment; Cell Densification
\end{IEEEkeywords}

\section{Introduction}
\label{sec:intro}

5G is the dominant technology that is going to revolutionize the services provided through cellular networks in the coming years \cite{obiodu20175g}. Clearly, in order to support the diverse services offered by this technology, new 5G Base Stations (BSs) have to be installed across the territory. In this context, a common opinion among the population is that living in proximity to a 5G BS is dangerous for health \cite{cousin2010public}. Although previous works in the literature have not shown any evidence of health effects triggered by living in proximity to radio BSs operating below maximum exposure limits (see e.g., \cite{whonote,chiaraviglio5g}), the debate about the installation of new 5G BSs is currently a hot topic, with several municipalities that are even denying the authorizations to install 5G BSs \cite{geneve}, on the basis of a precautionary principle. 

In this scenario, the concern that 5G will bring an uncontrolled proliferation of 5G BSs and consequently of Radio-Frequency ``Pollution'' (RFP) clearly emerges.\footnote{In the context of radio-frequency devices, the general public typically adopts the negative word ``pollution''. On the other hand, the scientific community generally exploits more neutral terms, e.g., exposure, radiation, and/or received power. In this work, we adopt the term ``pollution'' inside quotation marks in order to: i) improve the readability of this work also to the general public, ii) stress the fact that more neutral terms are used by the research community.}  However, is this anxiety corroborated by scientific evidence? Even if the scientific community well knows that this is not the case, to the best of our knowledge, no previous papers have tackled the impact of 5G BSs proliferation in terms of RFP by adopting models easily understood by non-technical readers. To shed light on this aspect, we introduce a set of simplifying (but worst case) assumptions, which lead us to derive a very simple model that evaluates the RFP that might be generated by a set of candidate 5G network deployments. We focus on regular BS coverage layouts, in which the coverage area of each BS is modeled with a regular shape (e.g., highway, square, hexagonal). Actually, such deployments are meaningful in dense population areas, like urban canyons and shopping malls, where the selection of BS sites is mainly driven by network capacity requirements. We then link together the coverage layout, the operating frequency, the radiated power, the minimum User Equipment (UE) sensitivity threshold, the level of RFP introduced by neighboring BSs and the channel propagation parameters. In this way, we obtain closed form expressions of RFP in terms of power received by a user at selected distances from the serving BS (e.g., at an average distance or a fixed one). By comparing the outcomes of different deployment types in a set of meaningful scenarios, we are able to assess the RFP variation, due to the adoption of one deployment w.r.t. another one. Our results indicate that the composite RFP tends to promptly decrease in intensity at the selected locations when the number of 5G BSs is increased. Moreover, the RFP contribution from neighboring BSs does not significantly impact the outcomes. Eventually, we show that, under specific circumstances (e.g., when the UE minimum sensitivity threshold is increased), there may be a slight increase of RFP from 5G BSs. 

Previous works in the literature cover other aspects related to 5G deployments, e.g., in terms of 5G techno-economic assessment \cite{oughton2019open}, joint power and ElectroMagnetic Fields (EMF) reduction \cite{matalatala2019multi}, and Specific Absorption Rate (SAR) evaluation \cite{sar}. We believe that all these aspects are surely of interest, but relatively orthogonal w.r.t. our work, which is instead tailored to the question on how 5G deployments may impact the RFP.


The rest of the paper is organized as follows. Sec.~\ref{sec:system_model} details the system model. Sec.~\ref{sec:scenarios} describes the scenarios. Sec.~\ref{sec:results} presents the results. Finally, Sec.~\ref{sec:conclusions} summarizes our outcomes and sketches future research activities.

\section{System Model}
\label{sec:system_model}

In this work, we consider the deployment of a set of 5G BSs to cover a set of geographic pixels in a scenario where a single operator provides a 5G service. Our analysis leverages some standard topological/regularity/propagation assumptions, namely: i) the BSs are placed on a regular layout, as we consider a dense deployment with a uniform distribution of users; consequently, each BS serves a portion of the total territory under consideration, ii) the BSs are characterized by common features in terms of coverage shape, coverage size, and adopted frequency; i.e., the same BS equipment is used across the set; iii) the propagation conditions are the same among the BSs in the set; e.g., the reliable coverage distance is sufficiently short to avoid modifications of the propagation model due to changes in the path loss exponent \cite{rappaport2017overview}.

\textbf{Key Assumptions.} We then introduce the following key assumptions, namely: i) the BS radiation pattern is omnidirectional, ii) the power of the BS is set to provide adequate reception quality imposed by a minimum UE sensitivity at the BS coverage edge,\footnote{The power of the BS is assumed to ensure the maximum limits imposed by law.  The investigation of the setting of BS power in the presence of multiple operators and pre-5G technologies operating over the territory is left for future work. This step clearly includes also alternative policies to set the BS power, different from a minimum sensitivity constraint considered in this work.} and iii) the level of RFP produced by the neighboring BSs is evaluated at the BS edge. In the following, we provide more details about why these assumptions are meaningful for our analysis.

Focusing on the first assumption, a real 5G BS generally exhibits a radiation pattern different than a omnidirectional one, because: i) sectorization is in general exploited, and ii) the extensive adoption of beamforming allows to concentrate the transmitted signal strength on specific locations. With sectorization, the radiation patterns match the orientation of the sectors. With beamforming, the actual RFP level that is received over the territory generally varies both in time and space, and it is normally estimated through statistical models, which demonstrate that the average RFP at a given pixel is substantially lower than the theoretically maximum value \cite{timeaveragedpower}. Assuming an omnidirectional radiation is a worst case scenario, in which: i) each pixel is served by a beam (i.e., the beams are activated simultaneously in all the directions), ii) each pixel is not affected by sectorization (i.e., for a given pixel to BS distance, the UE received power is constant across the entire geographic extent of the sector, even for pixels along the sector edge). This assumption leads to an over-estimation of the received RFP, thus substantiating our results. 

The second assumption is about the setting of the BS power. Let us denote with $\mathcal{I}$ and $\mathcal{P}$ the set of 5G BSs and the set of pixels, respectively. $P^E_{(i)}$ is the power emitted by BS $i \in \mathcal{I}$ to provide coverage within the pixels in its coverage area. $P^R_{\text{TH}}$ is the minimum power that has to be received by a pixel from a 5G BS providing adequate 5G coverage. For every pixel $p \in \mathcal{P}$ in the coverage area of BS $i$, a minimum sensitivity threshold has to be ensured through the following constraint:
\begin{equation}
\label{min_sens}
\frac{P^E_{(i)}}{d_{(p,i)}^\gamma \cdot f^\eta \cdot c} \geq P^R_{\text{TH}}
\end{equation}
where $d_{(p,i)}$ is the distance between pixel $p$ and 5G BS $i$, $\gamma$ is the path loss exponent for the distance, $f$ is the operating frequency, $\eta$ is the path loss exponent for the frequency, and $c$ is a baseline path loss. Since we consider a regular layout and a uniform user distribution, all the BSs belonging to the set radiate the same power, i.e., $P^E_{(i)}=P^E$. In addition, it is trivial to note that constraint (\ref{min_sens}) can be satisfied by setting $P^E$ as:
\begin{equation}
\label{eq:pe_comp}
P^E = P^R_{\text{TH}} \cdot d_{\text{MAX}}^\gamma \cdot f^\eta \cdot c
\end{equation}
where $d_{\text{MAX}}$ is the maximum coverage distance of a 5G BS. In other words, the received power at the BS edge is our reference to setup the BS power. In this scenario, the effect of interference from neighboring BSs and the impact of noise can be easily compensated by varying the values of $P^R_{\text{TH}}$. 

The third assumption involves the combined RFP computation due to the serving BS $s \in \mathcal{I}$ \textit{and} the neighboring ones $i \in \mathcal{I}^{\text{NEIGH}}, i \neq s$, where $\mathcal{I}^{\text{NEIGH}} \subset \mathcal{I}$ is the subset of neighboring BS whose RFP contribution can be sensed at pixel $p$. In general, the RFP $P^R_{(p)}$ that is received by a given pixel $p$ is expressed as:
\begin{equation}
\label{eq:gen_model}
P^R_{(p)}=\underbrace{\frac{P^E}{d_{(p,s)}^\gamma \cdot f^\eta \cdot c}}_\text{RFP from serving BS} + \underbrace{\sum_{i \in \mathcal \mathcal{I}^{\text{NEIGH}}} \frac{P^E}{d_{(p,i)}^\gamma \cdot f^\eta \cdot c}}_\text{RFP from neighboring BSs}
\end{equation}
Let us introduce the inter-site distance $d_{\text{SITE}}=2 \cdot \zeta \cdot d_{\text{MAX}}$, where $\zeta \in (0, 1)$ is a parameter set in order to avoid coverage holes. We then assume that the RFP of the neighbors is evaluated at distance $d_{\text{SITE}}/2=\zeta \cdot d_{\text{MAX}}$ for all neighbors $N^I=|\mathcal{I}^{\text{NEIGH}}|$. It is easy to note that this assumption leads to an Upper Bound (UB) of $P^R_{(p)}$ if  $d_{(p,s)}\leq \zeta \cdot d_{\text{MAX}}$, where $d_{(p,s)}$ is the distance between pixel $p$ and serving BS $s$. More formally, we have:
\begin{equation}
\label{eq:gen_model_2}
P^R_{(p)} \leq \underbrace{\frac{P^E}{d_{(p,s)}^\gamma \cdot f^\eta \cdot c}}_\text{RFP from serving BS} + \underbrace{N^I \cdot \frac{P^E}{\zeta^\gamma \cdot d_{\text{MAX}}^\gamma \cdot f^\eta \cdot c}}_\text{RFP from neighboring BSs (UB)}
\end{equation}
Although this assumption may appear too conservative at the first glance, as the distance between the current pixel $p$ and each neighbor is in general larger than $\zeta \cdot d_{\text{MAX}}$, in this work we show that the total RFP introduced by $N^I$ neighbors does not significantly affect the results, compared to the case in which no RFP contribution from the neighbors is assumed.

\begin{figure}[t]
\centering
\resizebox{0.8\columnwidth}{!}{
  \begin{tikzpicture}[scale=1.0,>=latex]
	\begin{scope}
	[
		every node/.style={
		regular polygon, 
		regular polygon sides=6,
		draw,
		minimum width=0.95cm,
		line width=2pt,
		fill=white,
		outer sep=0,
		inner sep=0}
	]
	\definecolor{lightsalmon}{RGB}{222,222,222};
	\definecolor{gray}{RGB}{205,205,205};
        \definecolor{lightgray}{RGB}{240,240,240};
         \definecolor{linen}{RGB}{250,240,230};
  \definecolor{coral}{RGB}{255,127,80};
 
   \node[dummy] (dummy) {};
   

    \node[minimum width=5cm] (2)  {};
    
    \node[circle,draw, align=center, fill=black, minimum width=0.1cm] (center2) at (2)  {};
    
    \node[minimum width=5cm, fill=white] (4) at (2.corner 4)[anchor=corner 2] {};
    
     \node[circle,draw, align=center, fill=black, minimum width=0.1cm] (center4) at (4)  {};
     

    \node[anchor=corner 3, minimum width=5cm] (3) at (4.corner 1) {};
    
     \node[circle,draw, align=center, fill=black, minimum width=0.1cm] (center3) at (3)  {};

    \node[anchor=corner 2, minimum width=5cm] (6) at (4.corner 4) {};
    
     \node[circle,draw, align=center, fill=black, minimum width=0.1cm] (center6) at (6)  {};

   \node[anchor=corner 5, minimum width=5cm] (1) at (4.corner 3) {};
   
    \node[circle,draw, align=center, fill=black, minimum width=0.1cm] (center1) at (1)  {};

   \node[anchor=corner 2, minimum width=5cm] (5) at (4.corner 6) {};
    \node[circle,draw, align=center, fill=black, minimum width=0.1cm] (center5) at (5)  {};

   \node[anchor=corner 1,minimum width=5cm] (7) at (4.corner 3) {};
    \node[circle,draw, align=center, fill=black, minimum width=0.1cm] (center7) at (7)  {};
    
    \draw[color=blue, line width=2pt,arrow data={0.1}{stealth},arrow data={0.3}{stealth},arrow data={0.5}{stealth}] (center2) -- (center4);
    \draw[color=blue, line width=2pt,arrow data={0.1}{stealth},arrow data={0.3}{stealth},arrow data={0.5}{stealth}] (center3) -- (center4);
    \draw[color=blue, line width=2pt,arrow data={0.1}{stealth},arrow data={0.3}{stealth},arrow data={0.5}{stealth}] (center1) -- (center4);
    \draw[color=blue, line width=2pt,arrow data={0.1}{stealth},arrow data={0.3}{stealth},arrow data={0.5}{stealth}] (center5) -- (center4);
    \draw[color=blue, line width=2pt,arrow data={0.1}{stealth},arrow data={0.3}{stealth},arrow data={0.5}{stealth}] (center6) -- (center4);
    \draw[color=blue, line width=2pt,arrow data={0.1}{stealth},arrow data={0.3}{stealth},arrow data={0.5}{stealth}] (center7) -- (center4);
    
     \node[minimum width=5cm, fill=white] (4bis) at (2.corner 4)[anchor=corner 2] {};
     
     \node[draw=none,minimum width=3.04cm, fill=none,dashed] (4avg) at (4) {};
     
     \node[circle,minimum width=3.04cm, fill=none,dashed] (4avgbis) at (4) {};
   
     \node[circle,draw, align=center, fill=black, minimum width=0.1cm] (center4bis) at (4)  {};
     
   \draw[color=blue, line width=2pt, dashed,arrow data={0.25}{stealth},arrow data={0.5}{stealth},arrow data={0.75}{stealth}] (center4) -- node[draw=none,fill=none,color=black,yshift=0.4cm,xshift=0.2cm] {\large $d_{\text{AVG}}$} (4avg.corner 3);
   
     \node[draw=none,fill=none,color=black,yshift=-0.7cm,xshift=0.95cm] at (center2) {\large $\zeta \cdot d_{\text{MAX}}$};
     
    
     \draw[color=black, line width=2pt] (center2) -- node[draw=none,fill=none,color=black,xshift=0.8cm] {\large $d_{\text{MAX}}$} (2.corner 2);

    \node[draw=none,fill=none,color=black,yshift=-1.9cm] at (center4) {\large Serving BS};
    
     \node[draw=none,fill=none,color=black,yshift=-1.5cm] at (center6) {\large Neighboring BS};

    \draw[line width=2pt,dotted] (center6) -- node[draw=none,fill=none,color=black,xshift=-1.9cm] {\large $d_{\text{SITE}}=2 \cdot \zeta \cdot d_{\text{MAX}}$} (center7);
  \end{scope}
	\end{tikzpicture}
}
\vspace{-0.7cm}
\caption{Graphical sketch of the distances appearing in the RFP computation of Eq.~(\ref{eq:gen_model_avg_rel}). The RFP of the serving BS is evaluated at $d_{\text{AVG}}$. The RFP level from neighboring BSs is evaluated at $\zeta \cdot d_{\text{MAX}}$.}
\label{fig:pollution_level}
\vspace{-5mm}
\end{figure}
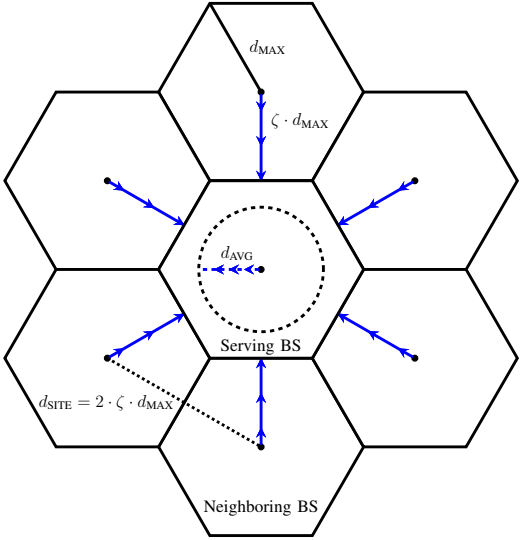

\textbf{Single Deployment RFP.}
Let us now consider a pixel located at an average distance $d_{\text{AVG}}$ from the serving BS. Clearly, it holds that $d_{\text{AVG}}<d_{\text{MAX}}$.\footnote{For regular deployments, $d_{\text{AVG}}<\zeta \cdot d_{\text{MAX}}$ also holds.}  By exploiting the right-hand side of Eq.~(\ref{eq:gen_model_2}), the RFP $P^R_{\text{AVG}}$ at average distance is expressed as:
\begin{equation}
\label{eq:gen_model_avg_rel}
P^R_{\text{AVG}}=\frac{P^E}{d_{\text{AVG}}^\gamma \cdot f^\eta \cdot c} + N^I \frac{P^E}{\zeta^\gamma \cdot d_{\text{MAX}}^\gamma \cdot f^\eta \cdot c}
\end{equation}
 Fig.~\ref{fig:pollution_level} illustrates the distances appearing in Eq.~\ref{eq:gen_model_avg_rel}, by considering a hexagonal layout. Specifically, the central BS is the serving one, while the adjacent BSs are the neighbors. Consequently, $N^I=6$ in this example. By analyzing the figure in detail, we observe that the RFP evaluation of the serving BS is done at distance $d_{\text{AVG}}$ while the RFP of the neighbors is performed at distance $\zeta \cdot d_{\text{MAX}}$. 

Since we adopt a regular shape for the coverage area, the average distance $d_{\text{AVG}}$ is expressed as $d_{\text{AVG}}=\alpha \cdot d_{\text{MAX}}$, where $\alpha \in (0,1)$ is a fixed parameter depending on the chosen coverage layout. Consequently, Eq.~(\ref{eq:gen_model_avg_rel}) can be rewritten as:
\begin{equation}
\label{eq:gen_model_avg_rel_2}
P^R_{\text{AVG}}=\frac{P^E \cdot \left[\alpha^{-\gamma}+N^I \cdot \zeta^{-\gamma}\right]}{d_{\text{MAX}}^\gamma \cdot f^\eta \cdot c} 
\end{equation}
By expressing $P^E$ as in Eq.~(\ref{eq:pe_comp}), we can simplify $P^R_{\text{AVG}}$ as:
\begin{equation}
\label{eq:gen_model_avg_rel_3}
P^R_{\text{AVG}}=P^R_{\text{TH}} \cdot \left[\alpha^{-\gamma}+N^I \cdot \zeta^{-\gamma}\right]
\end{equation}

Let us now consider a pixel at a fixed distance $d_{\text{FX}}$ from the serving BS. The relationship between $d_{\text{FX}}$ and $d_{\text{MAX}}$ is expressed as $d_{\text{FX}}=\beta \cdot d_{\text{MAX}}$, where $\beta \in (0,1)$ is a fixed parameter. By adopting a procedure similar to Eq.~(\ref{eq:gen_model_avg_rel})-(\ref{eq:gen_model_avg_rel_3}), we can formally introduce the RFP $P^R_{\text{FX}}$ at distance $d_{\text{FX}}$:
\begin{equation}
\label{eq:gen_model_min_rel_3}
P^R_{\text{FX}}=P^R_{\text{TH}} \cdot \left[\beta^{-\gamma}+N^I \cdot \zeta^{-\gamma} \right]
\end{equation}
In our scenarios, we will consider values of $d_{\text{FX}}$ in proximity to the 5G BSs, since in general people working/living close to the BSs are the ones expressing the highest concerns about RFP. However, the same model can be adopted also for distances up to $\zeta \cdot d_{\text{MAX}}$.

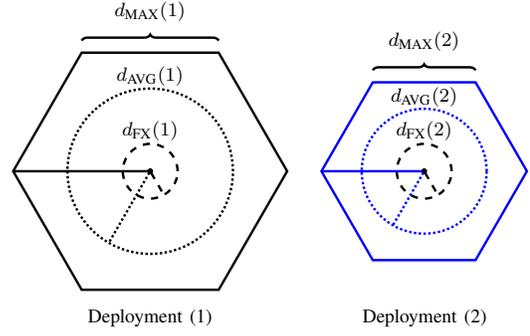
\begin{figure}[t]
\centering
\resizebox{0.8\columnwidth}{!}{
  \begin{tikzpicture}[scale=1.0,>=latex]
	\begin{scope}
	[
		every node/.style={
		regular polygon, 
		regular polygon sides=6,
		draw,
		minimum width=0.95cm,
		line width=1pt,
		outer sep=0,
		inner sep=0}
	]
	\definecolor{lightsalmon}{RGB}{222,222,222};
	\definecolor{gray}{RGB}{205,205,205};
        \definecolor{lightgray}{RGB}{240,240,240};
         \definecolor{linen}{RGB}{250,240,230};
  \definecolor{coral}{RGB}{255,127,80};
 
   \node[dummy] (dummy) {};
   

    \node[minimum width=4cm] (bighex)  {};
    
     \node[draw=none,minimum width=2.416cm, densely dotted] (davg)  {};

    \node[circle,minimum width=2.416cm, densely dotted] (davgbis)  {};
     
    \node[rectangle,draw=none,align=center,fill=none] (1text)  at ([yshift=0.6cm]bighex.north) {\footnotesize  $d_{\text{MAX}}(1)$};

    \draw[decorate,decoration={brace,raise=5pt},line width=1pt] (bighex.corner 2) -- (bighex.corner 1);
    
    \node[draw=none,minimum width=0.8cm, dashed] (dfx) {};

    \node[circle,minimum width=0.8cm, dashed] (dfxbis) {};
    
    \node[rectangle,draw=none,align=center,fill=none] (2text)  at ([yshift=0.35cm]davg.north) {\footnotesize $d_{\text{AVG}}(1)$};
    
    \node[rectangle,draw=none,align=center,fill=none] (3text)  at ([yshift=0.25cm]dfx.north) {\footnotesize  $d_{\text{FX}}(1)$};
    
    \node[circle,draw, align=center, fill=black, minimum width=0.05cm] (center) {};
    
    \draw[line width=1pt] (center) -- (bighex.west); 
    
    \draw[line width=1pt, densely dotted] (center) -- (davg.corner 4);
    
    \draw[line width=1pt, dashed] (center) -- (dfx.corner 5);
    
    \node[color=blue,minimum width=3cm] (smallhex) at ([xshift=2cm]bighex.east) {};

    \draw[decorate,decoration={brace,raise=5pt},line width=1pt] (smallhex.corner 2) -- (smallhex.corner 1);
    
     \node[rectangle,draw=none,align=center,fill=none] (4text)  at ([yshift=0.6cm]smallhex.north) {\footnotesize  $d_{\text{MAX}}(2)$};
    
     \node[draw=none,color=blue,minimum width=1.824cm, densely dotted] (davg2) at (smallhex)  {};

      \node[circle,color=blue,minimum width=1.824cm, densely dotted] (davg2bis) at (smallhex)  {};
     
     \node[draw=none,minimum width=0.8cm, dashed] (dfx2) at (smallhex)  {};

     \node[circle,minimum width=0.8cm, dashed] (dfx2bis) at (smallhex)  {};
     
     \node[rectangle,draw=none,align=center,fill=none] (5text)  at ([yshift=0.30cm]davg2.north) {\footnotesize  $d_{\text{AVG}}(2)$};
    
    \node[rectangle,draw=none,align=center,fill=none] (6text)  at ([yshift=0.25cm]dfx2.north) {\footnotesize  $d_{\text{FX}}(2)$};
    
    \node[circle,draw, align=center, fill=black, minimum width=0.05cm] (center2) at (smallhex) {};
    
    \draw[color=blue,line width=1pt] (center2) -- (smallhex.west);

    \draw[color=blue,line width=1pt, densely dotted] (center2) -- (davg2.corner 4);
    
    \draw[line width=1pt, dashed] (center2) -- (dfx2.corner 5);
    
    \node[rectangle,draw=none,align=center,fill=none] (7text)  at ([yshift=-1.8cm]dfx.south) {\footnotesize Deployment (1)};
    
    \node[rectangle,draw=none,align=center,fill=none] (8text)  at ([xshift=3.1cm]7text.east) {\footnotesize Deployment (2)};

  \end{scope}
	\end{tikzpicture}
}
\caption{An example of $d_{\text{MAX}}$, $d_{\text{AVG}}$ and $d_{\text{FX}}$ in two hexagonal deployments. $d_{\text{MAX}}(2)<d_{\text{MAX}}(1)$,  $d_{\text{AVG}}(2)=\alpha \cdot d_{\text{MAX}}(2)$, $d_{\text{AVG}}(1)=\alpha \cdot d_{\text{MAX}}(1)$. Consequently, it holds that $d_{\text{AVG}}(2)<d_{\text{AVG}}(1)$, while $d_{\text{FX}}(2)=d_{\text{FX}}(1)$ (figure best viewed in colors).}
\label{fig:hex_example}
\vspace{-5mm}
\end{figure}

Finally, Fig.~\ref{fig:hex_example} shows an example of $d_{\text{MAX}}$, $d_{\text{AVG}}$ and $d_{\text{FX}}$ in two hexagonal deployments, which are labelled as (1) and (2), respectively. Since $d_{\text{MAX}}(2)<d_{\text{MAX}}(1)$, it is easy to note that $d_{\text{AVG}}(2) < d_{\text{AVG}}(1)$. However, $d_{\text{FX}}(1)=d_{\text{FX}}(2)$.

\textbf{RFP Comparison among 5G Deployments.}
Let us now consider the comparison among two different 5G deployment types, denoted with indexes $(1)$ and $(2)$, respectively. Each deployment type is characterized by a given set of features, e.g., frequency $f$, propagation exponent $\gamma$, minimum power threshold $P^R_{\text{TH}}$, maximum distance $d_{\text{MAX}}$ and emitted (radiated) power ${P^E}$. We initially compare the deployments in terms of ratio of emitted powers $P^E(1)$ and $P^E(2)$, which is denoted as $\delta({P^E})$. By adopting Eq.~(\ref{eq:pe_comp}) to express $P^E(1)$ and $P^E(2)$, $\delta({P^E})$ becomes equal to:
\begin{equation}
\label{eq:ratio_power_e2}
\delta({P^E}) = \delta(d_{\text{MAX}})^{\gamma(1)} \cdot \frac{d_{\text{MAX}}(2)^{\gamma(1)}}{d_{\text{MAX}}(2)^{\gamma(2)}} \cdot \delta({P^R_{\text{TH}}}) \cdot \delta(f)^\eta \cdot \delta(c)
\end{equation}
where $\delta(d_{\text{MAX}})=\frac{d_{\text{MAX}}(1)}{d_{\text{MAX}}(2)}$, $\delta({P^R_{\text{TH}}})=\frac{P^R_{\text{TH}}(1)}{P^R_{\text{TH}}(2)}$, $\delta(f)=\frac{f(1)}{f(2)}$, $\delta(c)=\frac{c(1)}{c(2)}$. Clearly, when $\delta({P^E})>1$, the power radiated by deployment (1) is higher than the one of deployment (2). On the other hand, when $\delta({P^E})<1$, the opposite holds. Finally, when $\delta({P^E})=1$, there is no variation in the radiated power among the two deployments. 

\begin{table*}[t]
	\caption{5G Scenarios under consideration}
	\label{tab:scenarios}
	\scriptsize
\centering
    \begin{tabular}{|c|m{2.5cm}|m{3.0cm}|m{0.8cm}|m{3.5cm}|m{3.0cm}|m{0.6cm}|}
\hline
     \textbf{Scenario} & \textbf{Features} & $\delta(d_{\text{MAX}})$ & $\delta({P^R_{\text{TH}}})$ & $\frac{\gamma(1)}{\gamma(2)}$ & $\delta(f)$ & $\delta(c)$\\
\hline
S1 & Light Densification & 2 ($d_{\text{MAX}}(1)=500$~[m], $d_{\text{MAX}}(2)=250$~[m]) & 1 & 1 ($\gamma(1)=\gamma(2)=3)$ & 1 ($f_1=f_2 =700$~[MHz] & 1 \\
\hline
S2 & Moderate Densification & 5 ($d_{\text{MAX}}(1)=500$~[m], $d_{\text{MAX}}(2)=100$~[m]) & 1 & 1.43 ($\gamma(1)=3$, $\gamma(2)=2.1)$ & 1 ($f_1=f_2 =700$~[MHz] & 1\\
\hline
S3 & Light Densification, Frequency Change & 2 ($d_{\text{MAX}}(1)=500$~[m], $d_{\text{MAX}}(2)=250$~[m]) & 1 & 1 ($\gamma(1)=\gamma(2)=3)$ & 0.19 ($f_1=700$~[MHz], $f_2=3700$~[MHz]) & 1\\
\hline
S4 & Same Deployment, Service \& Frequency Change & 1 ($d_{\text{MAX}}(1)=500$~[m], $d_{\text{MAX}}(2)=500$~[m]) & 0.5 & 1 ($\gamma(1)=\gamma(2)=3)$ & 0.19 ($f_1=700$~[MHz], $f_2=3700$~[MHz])  & 1\\
\hline
S5 & Strong Densification, Service  \& Frequency Change &  10 ($d_{\text{MAX}}(1)=500$~[m], $d_{\text{MAX}}(2)=50$~[m]) & 0.5 & 1.43 ($\gamma(1)=3$, $\gamma(2)=2.1)$ & 0.19 ($f_1=700$~[MHz], $f_2=3700$~[MHz])  & 1\\
\hline
	\end{tabular}
\vspace{-3mm}
\end{table*}

\begin{figure*}[t]
\centering
\subfigure[S1,S3 - Deployment (1)]
{
	\includegraphics[width=0.465\columnwidth]{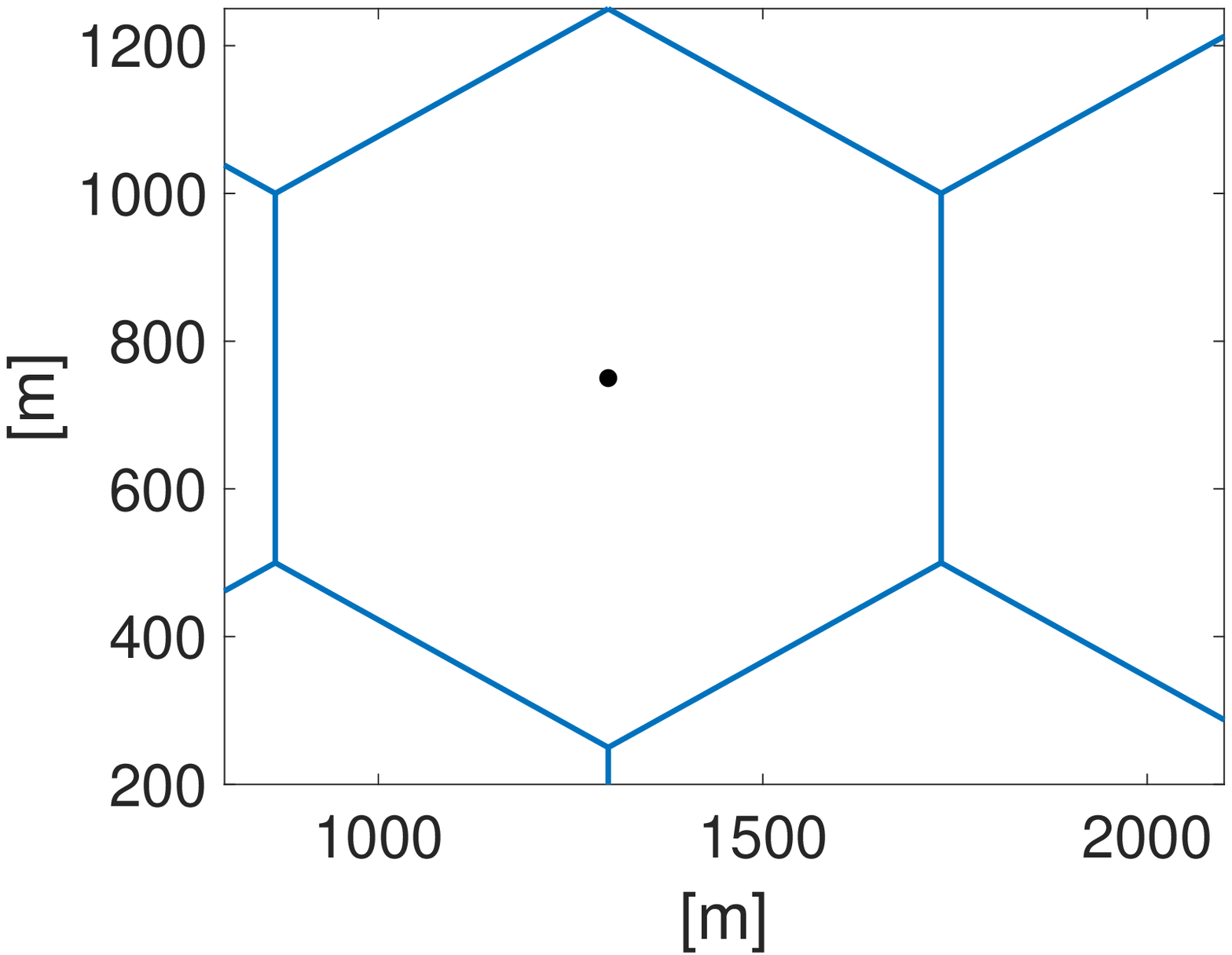}
	\label{fig:s1-1}
}
\subfigure[S1,S3 - Deployment (2)]
{
	\includegraphics[width=0.465\columnwidth]{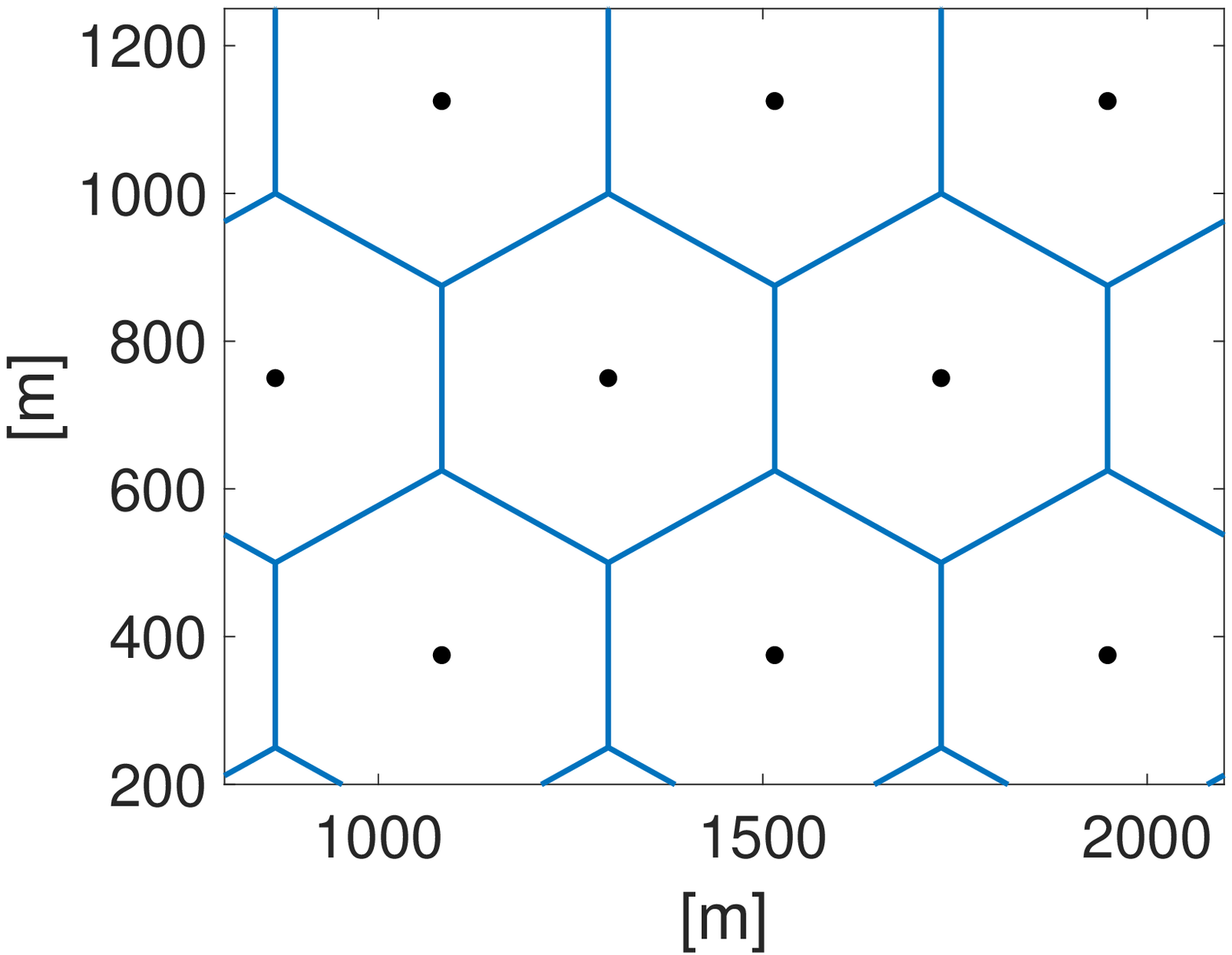}
	\label{fig:s1-2}
}
\subfigure[S2 - Deployment (1)]
{
	\includegraphics[width=0.465\columnwidth]{HEX_500}
	\label{fig:s2-1}
}
\subfigure[S2 - Deployment (2)]
{
	\includegraphics[width=0.465\columnwidth]{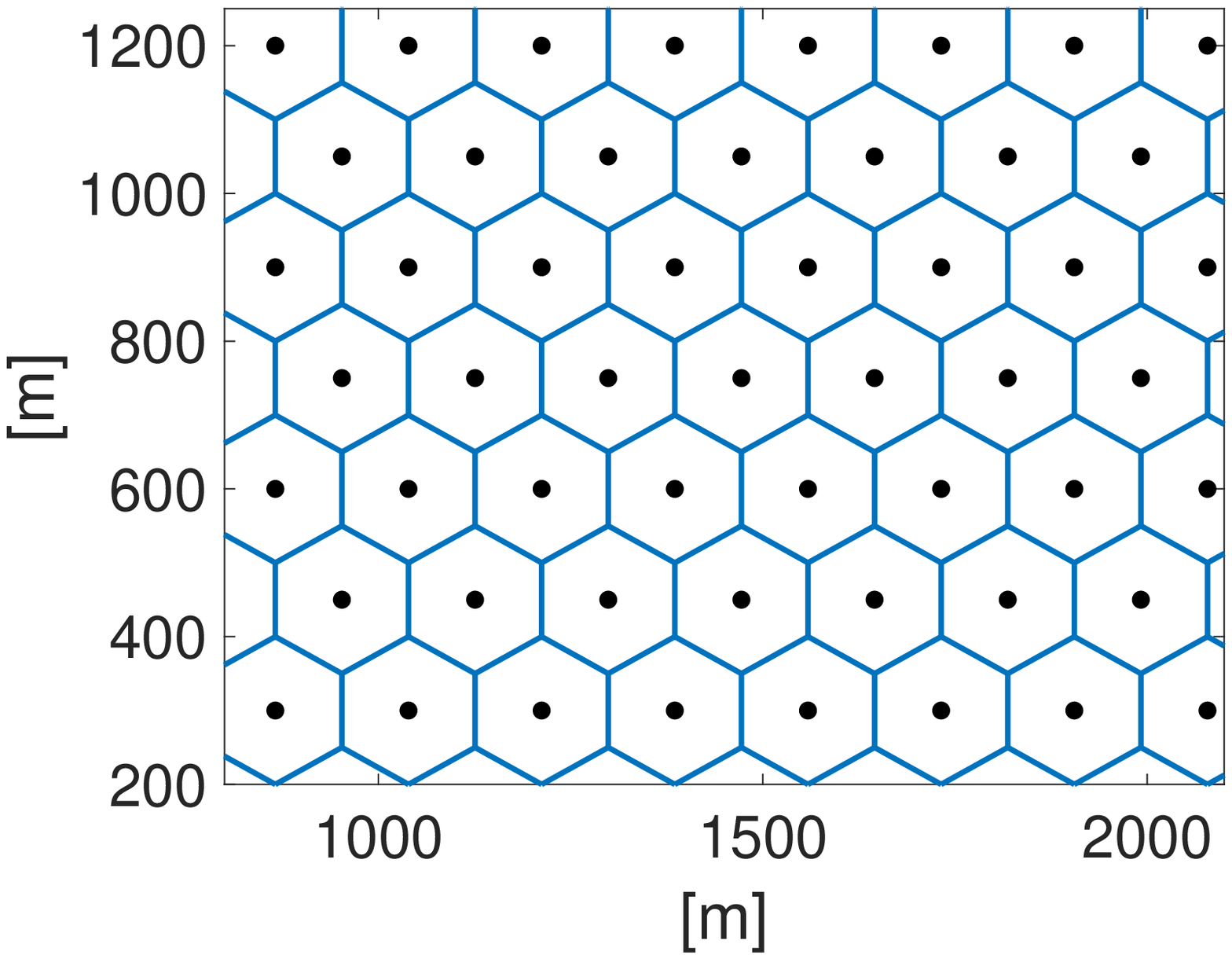}
	\label{fig:s2-2}
}

\subfigure[S4 - Deployment (1)]
{
	\includegraphics[width=0.465\columnwidth]{HEX_500}
	\label{fig:s4-1}
}
\subfigure[S4 - Deployment (2)]
{
	\includegraphics[width=0.465\columnwidth]{HEX_500}
	\label{fig:s4-2}
}
\subfigure[S5 - Deployment (1)]
{
	\includegraphics[width=0.465\columnwidth]{HEX_500}
	\label{fig:s5-1}
}
\subfigure[S5 - Deployment (2)]
{
	\includegraphics[width=0.465\columnwidth]{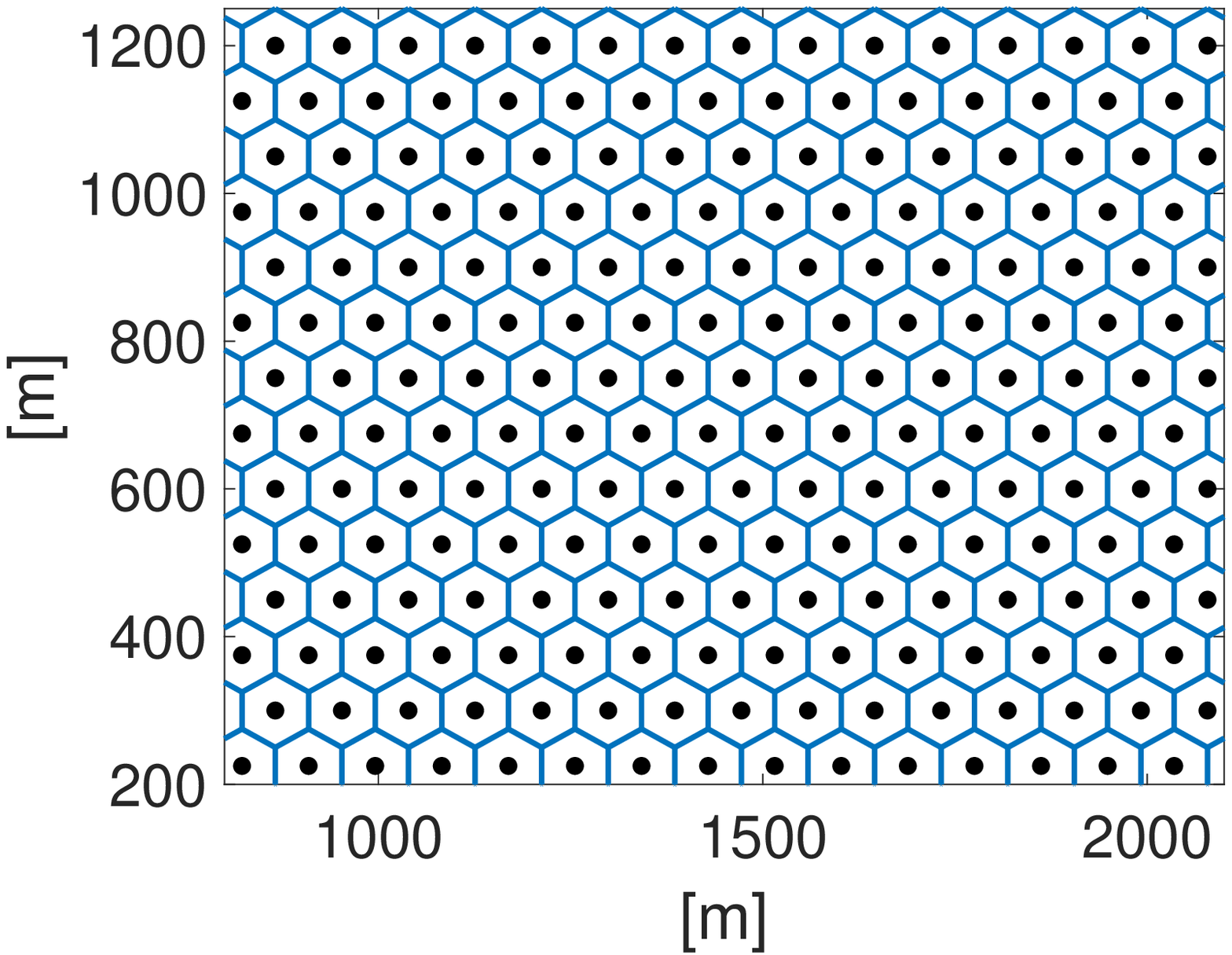}
	\label{fig:s5-2}
}

\caption{Visual representation of deployment (1) and deployment (2) across the scenarios S1-S5 (Hexagonal layout).}
\label{fig:s1-s5-hex}
\vspace{-5mm}
\end{figure*}

In the following, we compare the two deployments by introducing the RFP ratio at average distance, denoted as $\delta({P^R_{\text{AVG}}})$. By adopting Eq.~(\ref{eq:gen_model_avg_rel_3}), $\delta({P^R_{\text{AVG}}})$ is expressed as:
\begin{equation}
\label{eq:ratio_pr_avg}
\delta({P^R_{\text{AVG}}})=\delta({P^R_{\text{TH}}}) \cdot \frac{\alpha(1)^{-\gamma(1)}+N^I(1) \cdot \zeta^{-\gamma(1)}}{\alpha(2)^{-\gamma(2)}+N^I(2) \cdot \zeta^{-\gamma(2)}}
\end{equation} 
By assuming the same coverage layout in the two deployments, it holds that $\alpha(1)=\alpha(2)=\alpha$. Consequently, Eq.~(\ref{eq:ratio_pr_avg}) can be rewritten as:
\begin{equation}
\label{eq:ratio_pr_avg_2}
\delta({P^R_{\text{AVG}}})=\delta({P^R_{\text{TH}}})\cdot \frac{\alpha^{-\gamma(1)}+N^I(1) \cdot \zeta^{-\gamma(1)}}{\alpha^{-\gamma(2)}+N^I(2) \cdot \zeta^{-\gamma(2)}}
\end{equation}
Similarly to $\delta({P^E})$, $\delta({P^R_{\text{AVG}}})$ can take values $>1$, $<1$, or $=1$, depending on which deployment achieves the lowest RFP. However, differently from $\delta({P^E})$, $\delta({P^R_{\text{AVG}}})$ is the RFP received by a pixel at average distance $d_{\text{AVG}}(1) = \alpha \cdot d_{\text{MAX}}(1)$ in deployment (1) and average distance $d_{\text{AVG}}(2) = \alpha \cdot d_{\text{MAX}}(2)$ in deployment (2). Hence, $d_{\text{AVG}}(1) \neq d_{\text{AVG}}(2)$ if $d_{\text{MAX}}(1) \neq d_{\text{MAX}}(2)$. We believe that the metric $\delta({P^R_{\text{AVG}}})$ is meaningful, as it may be representative for a user living at an average distance in deployment (1) and at an average distance in deployment (2).

Finally, we introduce the RFP ratio at fixed distance, denoted as $\delta({P^R_{\text{FX}}})$. By adopting Eq.~(\ref{eq:gen_model_min_rel_3}), $\delta({P^R_{\text{FX}}})$ is expressed as:
\begin{equation}
\label{eq:ratio_power_min}
\delta({P^R_{\text{FX}}})=\delta({P^R_{\text{TH}}})\cdot \frac{\beta(1)^{-\gamma(1)}+N^I(1)\cdot \zeta^{-\gamma(1)}}{\beta(2)^{-\gamma(2)}+N^I(2)\cdot \zeta^{\gamma-(2)}}
\end{equation}

In this case, it is meaningful to compare the deployments at the same distance $d_{\text{FX}}(1)=d_{\text{FX}}(2)$. This setting is representative for a user living at the same distance from the serving BS in both the deployments. Hence, it holds that $\beta(2)=\beta(1) \cdot \delta(d_{\text{MAX}})$. Consequently, Eq.~(\ref{eq:ratio_power_min}) is rewritten as:
\begin{equation}
\label{eq:ratio_power_min_2}
\delta({P^R_{\text{FX}}})=\frac{\delta({P^R_{\text{TH}}}) \cdot [\beta(1)^{-\gamma(1)} +N^I(1) \cdot \zeta^{-\gamma(1)}]}{\beta(1)^{-\gamma(2)} \cdot \delta(d_{\text{MAX}})^{-\gamma(2)} +N^I(2)\cdot \zeta^{-\gamma(2)}}
\end{equation}  
Similarly to $\delta({P^R_{\text{AVG}}})$, also this metric can take values $>1$, $<1$, or $=1$. However, differently from $\delta({P^R_{\text{AVG}}})$, the distance between the user and the serving BS is kept constant in the two deployments. 

Summarizing, we compare deployment (1) and deployment (2) in terms of emitted power ratio $\delta({P^E})$ (Eq.~(\ref{eq:ratio_power_e2})), RFP ratio at average distance $\delta({P^R_{\text{AVG}}})$ (Eq.~(\ref{eq:ratio_pr_avg_2})), and RFP ratio at fixed distance $\delta({P^R_{\text{FX}}})$ (Eq.~(\ref{eq:ratio_power_min_2})).

\section{Scenarios}
\label{sec:scenarios}


We consider a set of representative scenarios, detailed in Tab.~\ref{tab:scenarios}. Each scenario includes a set of parameters to characterize the pair of deployments under consideration, namely $d_{\text{MAX}}$, $P_{\text{TH}}^R$, $\gamma$, $f$ and $c$. The numerical values for the propagation parameters $\gamma$ and $\eta$ are set in accordance to the 5G 3GPP CI propagation model detailed in \cite{rappaport2017overview} (3GPP TR 38.901 V14.0.0). By adopting this model, $\gamma=2.1$ and $\gamma=3$ for LOS and NLOS conditions, respectively. Moreover, we consider values of $d_{\text{MAX}}\leq500$~[m]: in this way, as reported by \cite{rappaport2017overview}, the exponent of the propagation model in LOS conditions does not change with distance. The frequency exponent $\eta$ is set to 2 for all the scenarios, as in \cite{rappaport2017overview}. In addition, we adopt the 5G Italian frequencies in the sub-6~[GHz] spectrum, which is the most promising option for offering coverage and a mixture of coverage and capacity. Eventually, the baseline path loss $c$ does not vary across the deployments. Actually, this term may include several factors, e.g., the fixed attenuation appearing in the Friis' law and the shadowing/scattering component \cite{rappaport2017overview}, but we prefer to keep it constant as it solely appears in the emitted power ratio $\delta({P^E})$, but not in the RFP ratios $\delta({P^R_{\text{AVG}}})$ and $\delta({P^R_{\text{FX}}})$. Consequently, $\delta(c)=1$ in all the scenarios.  Finally, each scenario is evaluated over different coverage layouts.
More precisely, we consider the following cases:
\begin{itemize}
\item Highway layout: the BSs are positioned on a strip, with $\zeta=1$ to avoid coverage holes;
\item Square layout: the BSs are positioned at the intersections of a Manhattan grid, with $\zeta=1/\sqrt{2}$;
\item Hexagonal layout: the BSs are placed on an hexagonal grid, with $\zeta=\sqrt{3}/2$.
\end{itemize}

Fig.~\ref{fig:s1-s5-hex} provides a visual representation of the two deployments under consideration in each scenario, for the hexagonal layout. In the following, we provide more details about each scenario:
\begin{itemize}
\item[S1)] light densification scenario (Fig.\ref{fig:s1-1}-\ref{fig:s1-2}): the only parameter (slightly) changing across deployment (1) and deployment (2) is $d_{\text{MAX}}$. In S1, deployment (2) is slightly denser than deployment (1), while all the other parameters do not vary across the two deployments;
\item[S2)]  moderate densification scenario (Fig.\ref{fig:s2-1}-\ref{fig:s2-2}), which is subject to a radical variation of $d_{\text{MAX}}$ and $\gamma$ across the two deployments. In S2, the operator adopts a denser deployment in (2) compared to (1). This choice is coupled with a different site deployment strategy and/or site configuration setting, which allows a better coverage over the territory. Consequently, $\gamma(2)<\gamma(1)$;
\item[S3)]  light densification plus a frequency change (Fig.\ref{fig:s1-1}-\ref{fig:s1-2}). In S3, both $d_{\text{MAX}}$ and $f$ are varied in the two deployments. Specifically, while the 700~[MHz] frequency in (1) is primary used to provide coverage, the 3700~[MHz] of (2) allows to achieve a good mixture of coverage and capacity. Moreover, we consider a slight reduction of  $d_{\text{MAX}}(2)$ compared to $d_{\text{MAX}}(1)$;
\item[S4)] variation of the operating frequency $f$ and of the minimum sensitivity threshold $P_{\text{TH}}^R$ (Fig.\ref{fig:s4-1}-\ref{fig:s4-2}). In S4, $d_{\text{MAX}}$ is not varied, while we assume an increase of $P_{\text{TH}}^R$ and $f$ when passing from deployment (1) to deployment (2). With these settings, the operator is able to support a 5G service demanding a higher amount of capacity in deployment (2) compared to (1);
\item[S5)] strong densification scenario (Fig.\ref{fig:s5-1}-\ref{fig:s5-2}), in which $d_{\text{MAX}}(2)$ is much lower than $d_{\text{MAX}}(1)$, $P_{\text{TH}}^R(2)>P_{\text{TH}}^R(1)$,  $f(2)>f(1)$, and $\gamma(2)<\gamma(1)$. In S5, the operator evaluates the impact of passing from a sparse set of 5G BSs to a very dense deployment. Clearly, this choice has an impact on the propagation conditions, as users in deployment (2) tend to be in LOS conditions w.r.t. the serving 5G BS, resulting in $\gamma(2)<\gamma(1)$. Moreover, the increase of the minimum sensitivity $P_{\text{TH}}^R$ and the adoption of an higher frequency $f$ in (2) compared to (1) allows the operator to provide a larger capacity to the users.
\end{itemize}

\begin{table}[t]
	\caption{Expression of $\alpha$ for different coverage layouts}
	\label{tab:alfa_expression}
	\scriptsize
\centering
    \begin{tabular}{|c|c|c|}
\hline
 & & \\[-0.8em]
    \textbf{Coverage Layout} & \textbf{Closed Formula} & \textbf{Numerical Value} \\ 
     & & \\[-0.8em]
    \hline
     & & \\[-0.8em]
    Highway & $\frac{1}{2}$ & 0.5 \\
    & & \\[-0.8em]
    \hline
     & & \\[-0.8em]
    Square & $\frac{\sqrt{2}}{6} (\sqrt{2} + \log(1+\sqrt{2})$ \cite{stone1991some} & 0.5411 \\
     & & \\[-0.8em]
    \hline
     & & \\[-0.8em]
    Hexagonal & $\left(\frac{1}{3} + \frac{\log{3}}{4}\right)$ \cite{stone1991some} & 0.6080 \\
     & & \\[-0.8em]
     \hline
      & & \\[-0.8em]
    Circle & $\frac{2}{3}$ \cite{larson1981urban} & 0.6667 \\
     & & \\[-0.1em]
     \hline
\end{tabular}
\vspace{-5mm}
\end{table}

We then set the amount of RFP $N^I$ generated by the neighboring BSs. To this point, we consider two distinct cases. In the first one, we assume perfect coverage provided by the BSs, i.e., each BS radiates power only over the covered area, without impacting areas covered by other BSs. Consequently, it holds that $N^I(1)=N^I(2)=0$. In the second case, we assume that each BS radiates power beyond the maximum distance $d_{\text{MAX}}$. In this way, the RFP of the current BS extends to the neighboring areas, which are covered by other BSs. Specifically, we assume an amount of RFP proportional to the number of adjacent BSs. Therefore, we set $N^I(1)=N^I(2)=2$, $N^I(1)=N^I(2)=8$, $N^I(1)=N^I(2)=6$ for the highway, square, and hexagonal layouts, respectively.

\begin{table*}[t]
	\caption{Closed-form expressions for $\delta({P^E})$, $\delta({P^R_{\text{AVG}}})$, $\delta({P^R_{\text{FX}}})$ in the different scenarios and for different values of $N^I$.}
	\label{tab:received_power_ratio}
	\scriptsize
\centering
    \begin{tabular}{|c|c|c|c|c|}
\hline
    & & & & \\[-0.8em]
    & \textbf{Scenario} & $\delta({P^E})$ & $\delta({P^R_{\text{AVG}}})$ & $\delta({P^R_{\text{FX}}})$\\
    & & & & \\[-0.8em]
    \cline{2-5}
    & & & & \\[-0.8em]
    & S1 & $\delta(d_{\text{MAX}})^{\gamma(1)}$ &  1 & $\delta(d_{\text{MAX}})^{\gamma(1)}$ \\
    & & & & \\[-0.8em]
\cline{2-5}
    & & & & \\[-0.8em]
   & S2 & $\delta(d_{\text{MAX}})^{\gamma(1)} \cdot d_{\text{MAX}}(2)^{\gamma(1)-\gamma(2)}$ &  $\alpha^{\gamma(2)-\gamma(1)} $ & $\beta(1)^{\gamma(2)-\gamma(1)} \cdot \delta(d_{\text{MAX}})^{\gamma(2)}$ \\
    & & & & \\[-0.8em]
\cline{2-5}
    & & & & \\[-0.8em]
  & S3 & $\delta(d_{\text{MAX}})^{\gamma(1)} \cdot \delta(f)^\eta$ &  1 & $\delta(d_{\text{MAX}})^{\gamma(1)}$ \\
    & & & & \\[-0.8em]
\cline{2-5}
    & & & & \\[-0.8em]
  & S4 & $ \delta({P^R_{\text{TH}}}) \cdot \delta(f)^\eta $  & $ \delta({P^R_{\text{TH}}})$ & $ \delta({P^R_{\text{TH}}})$ \\
    & & & & \\[-0.8em]
\cline{2-5}
    & & & & \\[-0.8em]
 \multirow{-9}{*}{\begin{sideways}$N^I(1)=N^I(2)=0$\end{sideways}} & S5 & $\delta(P^R_{\text{TH}}) \cdot \delta(d_{\text{MAX}})^{\gamma(1)} \cdot d_{\text{MAX}}(2)^{\gamma(1)-\gamma(2)} \cdot \delta(f)^\eta $  & $\alpha^{\gamma(2)-\gamma(1)} \cdot \delta({P^R_{\text{TH}}})$ & $\beta(1)^{\gamma(2)-\gamma(1)} \cdot \delta(d_{\text{MAX}})^{\gamma(2)} \cdot \delta({P^R_{\text{TH}}})$\\
    & & & & \\[-0.8em]
\hline
    & & & & \\[-0.8em]
  & S1 & $\delta(d_{\text{MAX}})^{\gamma(1)}$ &  1 & $\frac{ \beta(1)^{-\gamma(1)} +N^I(1) \cdot \zeta^{-\gamma(1)}}{\beta(1)^{-\gamma(1)} \cdot \delta(d_{\text{MAX}})^{-\gamma(1)} +N^I(1)\cdot \zeta^{-\gamma(1)}}$ \\
    & & & & \\[-0.8em]
\cline{2-5}
    & & & & \\[-0.8em]
  & S2 & $\delta(d_{\text{MAX}})^{\gamma(1)} \cdot d_{\text{MAX}}(2)^{\gamma(1)-\gamma(2)}$ &  $\frac{\alpha^{-\gamma(1)}+N^I(1) \cdot \zeta^{-\gamma(1)}}{\alpha^{-\gamma(2)}+N^I(1) \cdot \zeta^{-\gamma(2)}}$ & $\frac{ \beta(1)^{-\gamma(1)} +N^I(1) \cdot \zeta^{-\gamma(1)}}{\beta(1)^{-\gamma(2)} \cdot \delta(d_{\text{MAX}})^{-\gamma(2)} +N^I(1)\cdot \zeta^{-\gamma(2)}}$ \\
    & & & & \\[-0.8em]
\cline{2-5}
    & & & & \\[-0.8em]
 &  S3 & $\delta(d_{\text{MAX}})^{\gamma(1)} \cdot \delta(f)^\eta$ &  1 & \multicolumn{1}{c|}{$\frac{ \beta(1)^{-\gamma(1)} +N^I(1) \cdot \zeta^{-\gamma(1)}}{\beta(1)^{-\gamma(1)} \cdot \delta(d_{\text{MAX}})^{-\gamma(1)} +N^I(1)\cdot \zeta^{-\gamma(1)}}$} \\
    & & & & \\[-0.8em]
\cline{2-5}
    & & & & \\[-0.8em]
 &  S4 &  $ \delta({P^R_{\text{TH}}}) \cdot \delta(f)^\eta $ & $\delta({P^R_{\text{TH}}})$ & \multicolumn{1}{c|}{$\delta({P^R_{\text{TH}}})$} \\
    & & & & \\[-0.8em]
\cline{2-5}
    & & & & \\[-0.8em]
\multirow{-9}{*}{\begin{sideways}$N^I(1)=N^I(2)>0$\end{sideways}} &  S5 & $\delta(P^R_{\text{TH}}) \cdot \delta(d_{\text{MAX}})^{\gamma(1)} \cdot d_{\text{MAX}}(2)^{\gamma(1)-\gamma(2)} \cdot \delta(f)^\eta $  & $\delta({P^R_{\text{TH}}})\cdot \frac{\alpha^{-\gamma(1)}+N^I(1) \cdot \zeta^{-\gamma(1)}}{\alpha^{-\gamma(2)}+N^I(1) \cdot \zeta^{-\gamma(2)}}$ & $\frac{\delta({P^R_{\text{TH}}}) \cdot \left[ \beta(1)^{-\gamma(1)} +N^I(1) \cdot \zeta^{-\gamma(1)}\right]}{\beta(1)^{-\gamma(2)} \cdot \delta(d_{\text{MAX}})^{-\gamma(2)} +N^I(1)\cdot \zeta^{-\gamma(2)}}$\\
    & & & & \\[-0.1em]
\hline
	\end{tabular}
\vspace{-5mm}
\end{table*}

In the following, we set the values of $\beta(1)$, $\beta(2)$, $\alpha$ across the deployments. Focusing on $\beta(1)$, we initially assume that the RFP is evaluated in close proximity to the serving BS in deployment (1). Therefore, we set $\beta(1)=0.05$ in all the scenarios. Since $d_{\text{FX}}=\beta(1) \cdot d_{\text{MAX}}(1)$, this corresponds in assessing the RFP for a user at $d_{\text{FX}}=25$~[m] from the serving BS. Also, we recall that $\beta(2)$ is equal to $\beta(1) \cdot \delta(d_{\text{MAX}})$.  Focusing on $\alpha$,  Tab.~\ref{tab:alfa_expression} reports the closed-form expression and the numerical value for each coverage layout. For the highway layout, $\alpha=0.5$, as this value corresponds to the average distance in the interval $[0,1]$ for a BS centered in $x=0$ with $d_{\text{MAX}}=1$, i.e., $\alpha=\frac{1}{2} \int_{-1}^{1} |x| dx = 1/2$. For the square and hexagonal cases, the average distance is computed as:
\begin{equation}
\label{eq:avg_distance}
\alpha=\int_{-\infty}^{+\infty} \int_{-\infty}^{+\infty} f(x,y) \sqrt{(x^2+y^2)}\,dx\,dy
\end{equation} 
where $f(x,y)$ is the probability density function of a square/hexagon with $d_{\text{MAX}}=1$ centered in $(0,0)$. To solve Eq.~(\ref{eq:avg_distance}), we adopt the closed-form expressions of average distance retrieved by \cite{stone1991some}. We refer the interested reader to \cite{stone1991some} for the formal proofs about average distance computation in these two cases. Finally, Tab.~\ref{tab:alfa_expression} reports as a term of comparison the upper bound of $\alpha$, which is computed from a circle layout. Interestingly, we can note that the numerical value of $\alpha$ is increasing when passing from the square to the hexagonal layout, but still below the upper bound.

\section{Results}
\label{sec:results}

We initially compute the closed-form expressions of RFP metrics. We then provide a numerical evaluation to better quantify the impact in terms of potential RFP. 

\textbf{Closed-Form Expressions of RFP Metrics.} Tab.~\ref{tab:received_power_ratio} reports the closed-form expressions for $\delta({P^E})$, $\delta({P^R_{\text{AVG}}})$, $\delta({P^R_{\text{FX}}})$ over the different scenarios, by considering the two different neighboring RFP options ($N^I=0$ or $N^I>0$). The ratio of this step is in fact to provide a ready-to-use tool when considering specific settings for the parameters, e.g., maximum distance increase, frequency increase, change in propagation conditions, etc. 

Let us first consider the scenarios with $N^I=0$ (upper part of Tab.~\ref{tab:received_power_ratio}). In the light densification scenario (S1), we vary $d_{\text{MAX}}$ across the two deployments. Consequently, $\delta(d_{\text{MAX}})$ is the only term affecting the RFP. By recalling that $\delta(d_{\text{MAX}})>1$ in S1, we can observe that: i) $\delta({P^E})\gg1$, ii) $\delta({P^R_{\text{AVG}}})$ does not change across the two deployments, iii) $\delta({P^R_{\text{FX}}})\gg1$. Consequently, deployment (2) achieves a lower RFP level than deployment (1) at a fixed distance, i.e., a light densification is able to decrease the RFP that is measured at a fixed distance, independently from the chosen coverage layout. Moreover, there is no variation in terms of RFP at an average distance from the serving BS.\footnote{We recall that the fixed distance does not change among deployment (1) and deployment (2), while the average distance is equal to $\alpha \cdot d_{\text{MAX}}(1)$ and $\alpha \cdot d_{\text{MAX}}(2)$ for deployment (1) and deployment (2), respectively.} Eventually, the emitted power in deployment (2) is lower than the one in deployment (1).

Focusing on the moderate densification scenario (S2), both $d_{\text{MAX}}$ and $\gamma$ are varied. Since $\gamma(1)>\gamma(2)$ and $\delta(d_{\text{MAX}})>1$, it holds that $\delta({P^E})\gg1$. Moreover, $\alpha$ appears in the expression of $\delta({P^R_{\text{AVG}}})$ (see Tab.~\ref{tab:received_power_ratio}). By recalling that $\alpha<1$ (see Tab.~\ref{tab:alfa_expression}), we can observe that $\delta({P^R_{\text{AVG}}})>1$. Moreover, it is easy to note that $\delta({P^R_{\text{FX}}})\gg1$. Therefore, a moderate densification, coupled to an improvement of the channel conditions, is able to noticeably reduce the potential RFP at the selected locations.

We then move our attention to S3, i.e., the scenario with a light densification and a frequency change. In this case, it is necessary to consider the specific values set to $d_{\text{MAX}}$ and $f$ in order to derive the values of $\delta({P^E})$. However, it is interesting to see that both $\delta({P^R_{\text{AVG}}})$, $\delta({P^R_{\text{FX}}})$ do not depend on $\delta(f)$, and they are the same as S1. Consequently, we can conclude that also for S3 the light densification and the frequency change results in a reduction of the RFP at a fixed distance, while there is no RFP variation at an average distance. 

In S4, the main goal is to provide better service in deployment (2) w.r.t. deployment (1). In this case, the only parameters varied are $f$ and $P^R_{\text{TH}}$, resulting in $\delta(f)<1$ and $\delta(P^R_{\text{TH}})<1$. By inspecting the expression of $\delta({P^E})$ reported in Tab.~\ref{tab:received_power_ratio}, we can clearly see that the emitted power is increased in deployment (2) w.r.t. deployment (1). Moreover, since $\delta({P^R_{\text{AVG}}})$ and $\delta({P^R_{\text{FX}}})$ depend solely on $\delta(P^R_{\text{TH}})$ in this case, we can conclude that the RFP is potentially higher in deployment (2) compared to deployment (1). Importantly, however, the potential RFP increase can be controlled by properly tuning the minimum reception threshold ratio $\delta(P^R_{\text{TH}})$.

Eventually, we consider S5, i.e., the scenario with a strong densification, coupled with a service and frequency change. Since S5 is a mixture of the previous ones, the expressions of $\delta({P^E})$, $\delta({P^R_{\text{AVG}}})$, $\delta({P^R_{\text{FX}}})$ include the terms $\delta(P^R_{\text{TH}})$ and $\delta(f)$, which in this case are lower than unity. As a result, the values of the RFP metrics can not be easily inferred in advance, and they have to be numerically evaluated, by considering the whole set of input parameters.

In the following, we move our attention to the $N^I>0$ case (bottom part of Tab.~\ref{tab:received_power_ratio}). Clearly, the expression of $\delta({P^E})$ does not change w.r.t. the $N^I=0$ case, since this term is not affected by $N^I$.\footnote{The level of RFP from neighboring BSs may also have an impact on the level of interference experienced by a particular pixel. We plan to investigate this issue as a future work.} Focusing then on $\delta({P^R_{\text{AVG}}})$, a change in the expression only occurs for S2 and S5, i.e., when the path loss exponent varies across the deployments. On the other hand, S1, S3 and S4 are subject to the same expressions of $\delta({P^R_{\text{AVG}}})$ w.r.t. to the $N^I=0$ case. Therefore, the same comments hold. Finally, we consider $\delta({P^R_{\text{FX}}})$: apart from S4 (which is the same as in the $N^I=0$ case), all the other scenarios  require a numerical evaluation to assess the potential RFP impact.

\begin{figure}[t]
\centering
\subfigure[$\delta({P^R_{\text{AVG}}})$, $N^I=0$]
{
	\includegraphics[width=0.45\columnwidth]{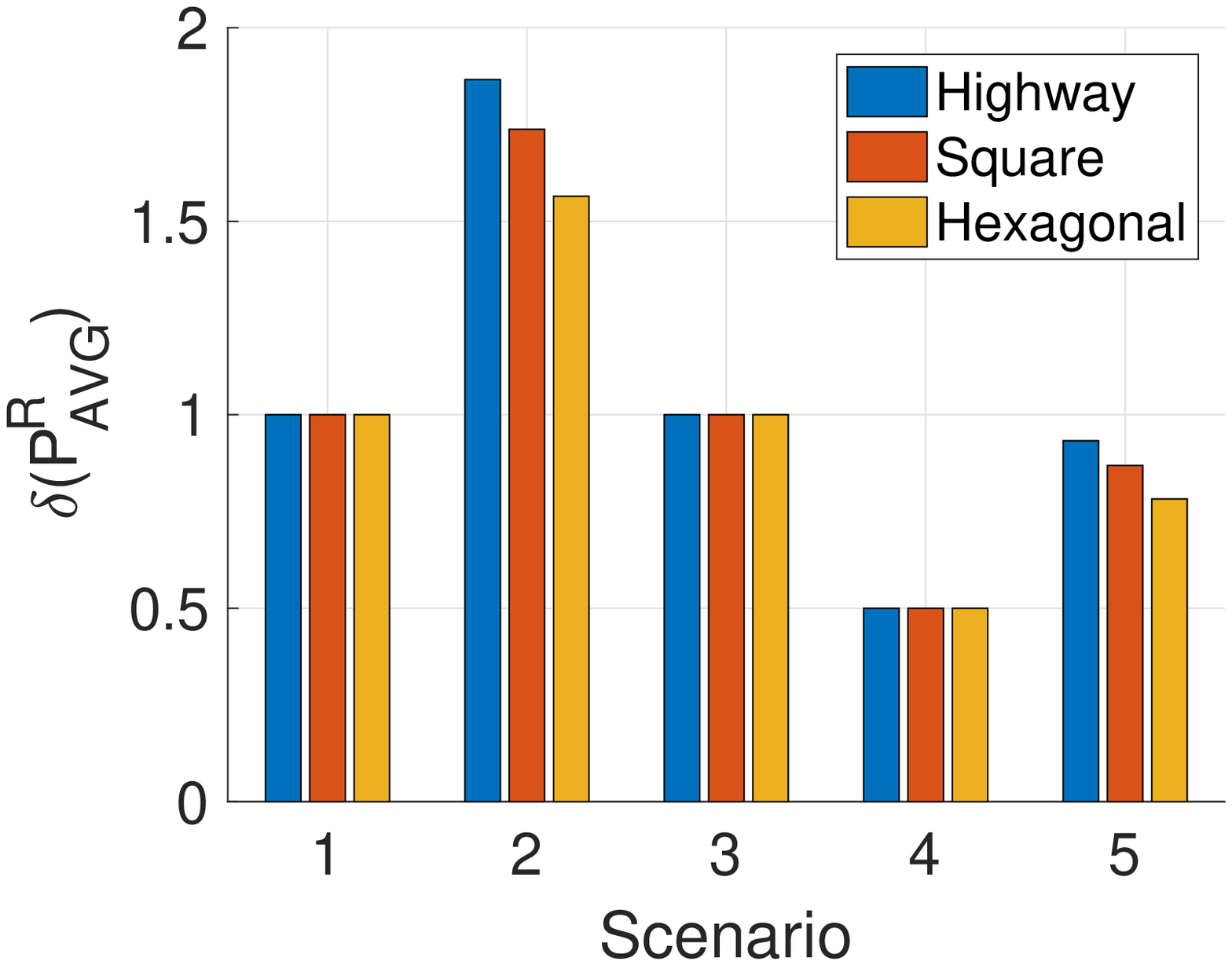}
	\label{fig:pr_avg_no_N_I}
}
\subfigure[$\delta({P^R_{\text{AVG}}})$, $N^I>0$]
{
	\includegraphics[width=0.45\columnwidth]{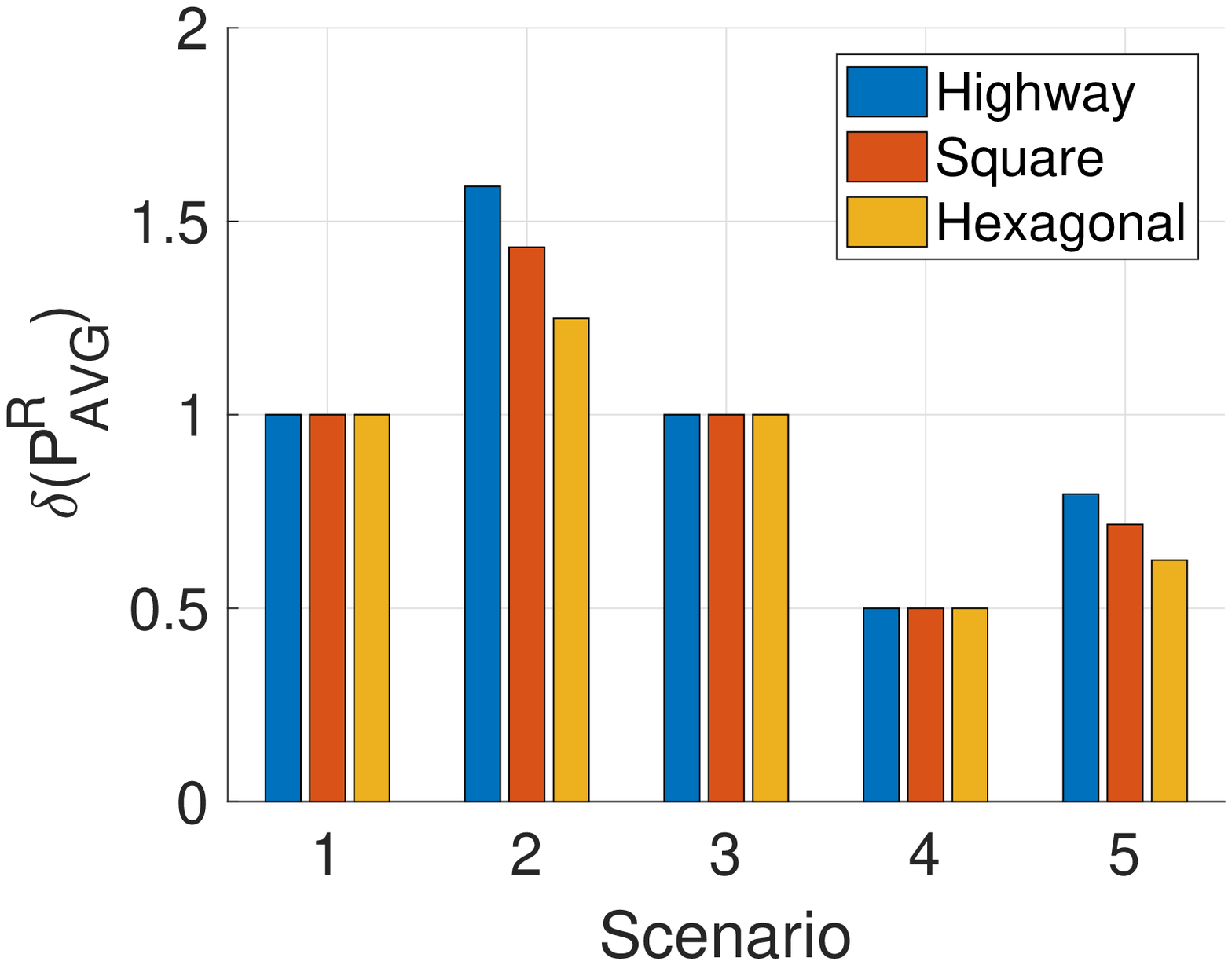}
	\label{fig:pr_avg_with_N_I}
}

\subfigure[$\delta({P^R_{\text{FX}}})$, $N^I=0$]
{
	\includegraphics[width=0.45\columnwidth]{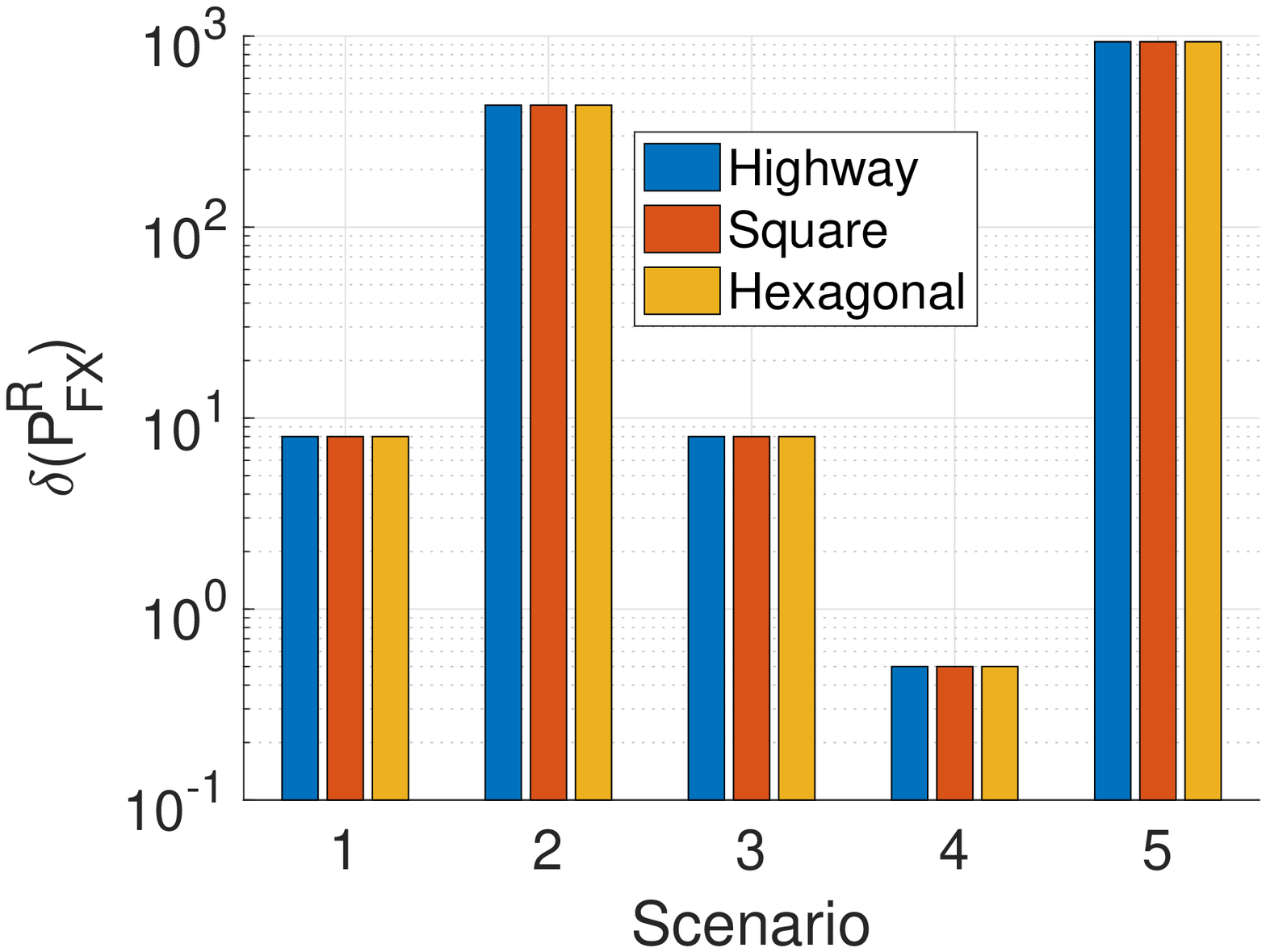}
	\label{fig:pr_max_no_N_I}
}
\subfigure[$\delta({P^R_{\text{FX}}})$, $N^I>0$]
{
	\includegraphics[width=0.45\columnwidth]{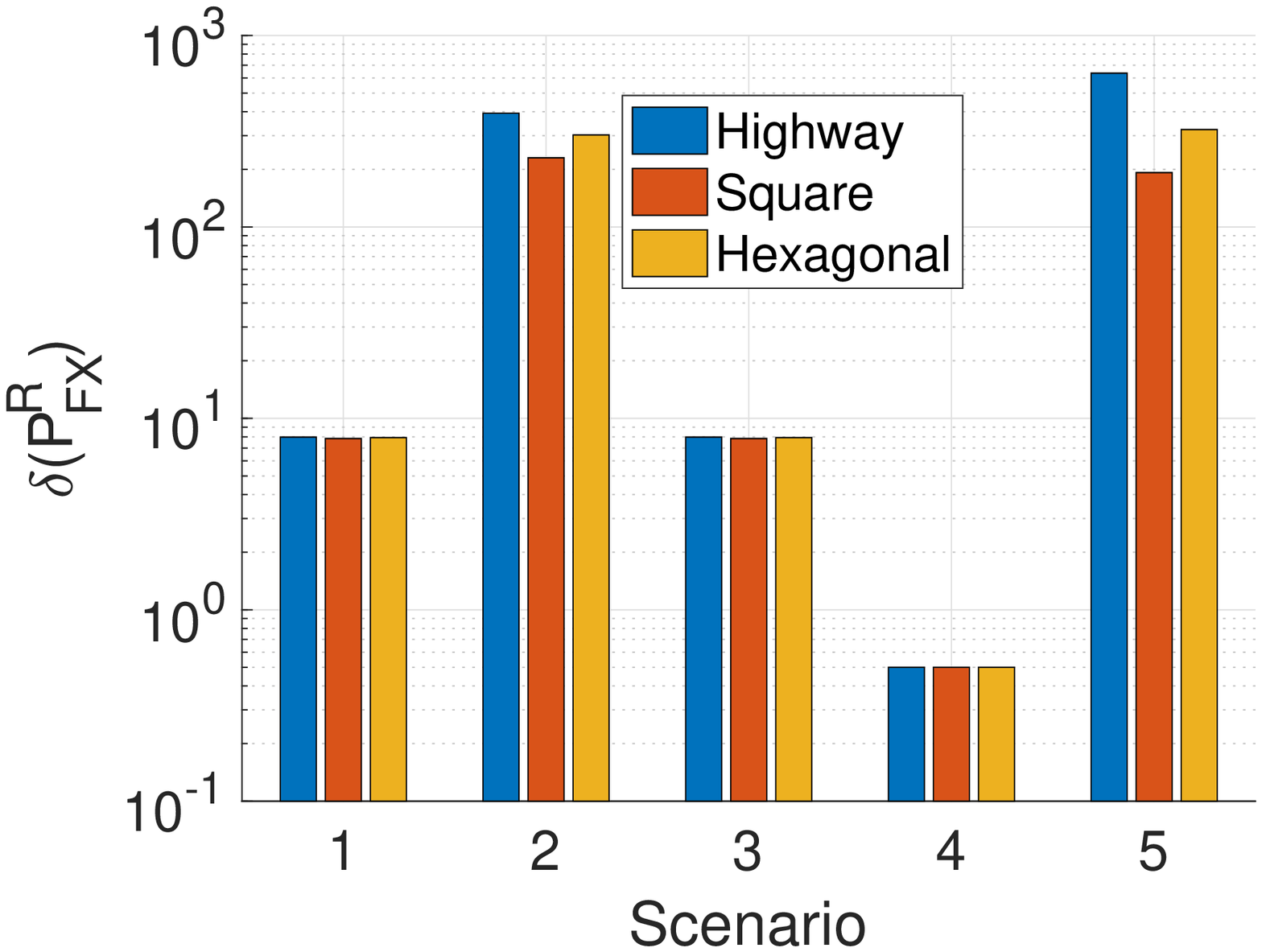}
	\label{fig:pr_max_with_N_I}
}
\caption{RFP ratio at average distance $\delta({P^R_{\text{AVG}}})$ and at fixed distance $\delta({P^R_{\text{FX}}})$ across the different scenarios, for different values of $N_I$.}
\label{fig:pr_avg_pr_max}
\vspace{-5mm}
\end{figure}

\textbf{Numerical Evaluation of RFP.} We then solve the expressions in Tab.~\ref{tab:received_power_ratio} by considering the whole set of parameters described in Sec.~\ref{sec:scenarios}, in order to compute $\delta({P^R_{\text{AVG}}})$ and $\delta({P^R_{\text{FX}}})$.\footnote{The numerical values of $\delta(P^E)$ are omitted due to the lack of space.} Fig.~\ref{fig:pr_avg_pr_max} illustrates the values of $\delta({P^R_{\text{AVG}}})$ and $\delta({P^R_{\text{FX}}})$ over the scenarios S1-S5, the highway/square/hexagonal layouts, and the two options of RFP from neighboring BSs (i.e., $N^I=0$ and $N^I>0$). Focusing first on $\delta({P^R_{\text{AVG}}})$ and the $N^I=0$ case (Fig.~\ref{fig:pr_avg_no_N_I}) we can note that the RFP ratio exceeds unity in scenario S2. Moreover, $\delta({P^R_{\text{AVG}}})$ increases when passing from the hexagonal to the highway layout. This is an expected result, since: i) we have already shown in Tab.~\ref{tab:received_power_ratio} that $\delta({P^R_{\text{AVG}}})=\alpha^{\gamma(2)-\gamma(1)}$, and hence this term is higher than one, ii) we have demonstrated in Tab.~\ref{tab:alfa_expression} that $\alpha$ is decreased when passing from the hexagonal to the highway layout, (iii) the lower is $\alpha$, the higher is $\delta({P^R_{\text{AVG}}})$. Focusing on S1 and S3, $\delta({P^R_{\text{AVG}}})=1$, independently from the coverage layout (as expected). Eventually, $\delta({P^R_{\text{AVG}}})=\delta({P^R_{\text{TH}}})=0.5$ in S4. Finally, $\delta({P^R_{\text{AVG}}})<1$ in S5, i.e., deployment (2) results in slightly higher RFP than deployment (1) when the evaluation is done at the average distance.

We then move our attention to $\delta({P^R_{\text{AVG}}})$ with $N^I>0$ (Fig.~\ref{fig:pr_avg_with_N_I}). Interestingly, the RFP ratio tends to be decreased in S2 and S5 compared to the $N^I=0$ case. Also, the highway layout always presents a better RFP ratio $\delta({P^R_{\text{AVG}}})$ compared to the square and hexagonal ones. This result is meaningful, as the RFP level that results from neighboring BSs is higher for the square/hexagonal layout compared to the highway scenario.

In the following, we consider the RFP ratio $\delta({P^R_{\text{FX}}})$ at fixed distance with $N^I=0$, shown in Fig.~\ref{fig:pr_max_no_N_I}. Interestingly, $\delta({P^R_{\text{FX}}})$ is clearly greater than unity in all the scenarios (except from S4). Specifically, the RFP ratio at fixed distance is close to 1000 for S5, higher than 400 for S2, and close to 8 for S1 and S3. These scenarios are all subject to a densification of 5G BSs (as $d_{\text{MAX}}(2)<d_{\text{MAX}}(1)$), which is beneficial for reducing the level of potential RFP at fixed distance. The only scenario with $\delta({P^R_{\text{FX}}})<1$ is S4, which we recall does not include any BS densification, but only a service and a frequency change.

Finally,  Fig.~\ref{fig:pr_max_with_N_I} reports the values of $\delta({P^R_{\text{FX}}})$ for the $N^I>0$ case. As expected, the level of RFP coming from the neighboring BSs tends to decrease $\delta({P^R_{\text{FX}}})$ in the S2 and S5 scenarios. However, $\delta({P^R_{\text{FX}}})$ is always higher than 100 in these scenarios, thus demonstrating that deployment (2) greatly reduces the RFP compared to deployment (1).

\section{Summary and Future Works}
\label{sec:conclusions}

Our analysis provided an objective approach to addressing the concern that a proliferation of 5G BSs will result into an uncontrollable RFP of the population. To this aim, we have derived a simple - yet effective - model to compare pairs of 5G deployments. 
Results demonstrate that the supposed increase of potential RFP due to the proliferation of 5G BSs is not supported by the evidence in most of the considered scenarios. In particular, when the number of BSs is increased (i.e., $d_{\text{MAX}}$ is reduced), the level of RFP at a fixed distance is decreased by almost three orders of magnitude. Ultimately, we have analyzed the conditions that may result a slight increase in RFP. 
Finally, the RFP level from neighboring BSs does not significantly affect the results in most of the scenarios.

We believe that this work can be an important first step to a more comprehensive approach, which may evaluate: i) the amount of RFP introduced in every location of the territory (and not only at selected locations), ii) the variation of power requirements over space and time (e.g. to match an increase or decrease in traffic demand), iii) the adoption of directional antennas, iv) the exploration of frequencies in the mm-Waves bands, and v) the impact of irregular coverage designs.

\section*{Acknowledgements}
This work has received funding from the H2020 Locus Project (grant agreement n. 871249).

\bibliographystyle{ieeetr}

\begin{thebibliography}{10}

\bibitem{obiodu20175g}
E.~Obiodu and M.~Giles, ``The 5G era: age of boundless connectivity and
  intelligent automation,'' {\em GSM Assocation}, 2017.

\bibitem{cousin2010public}
M.-E. Cousin and M.~Siegrist, ``The public's knowledge of mobile communication
  and its influence on base station siting preferences,'' {\em Health, Risk \&
  Society}, vol.~12, no.~3, pp.~231--250, 2010.

\bibitem{whonote}
{\em Electromagnetic fields and public health - Base stations and wireless
  technologies}.
\newblock Available at
  \url{https://www.who.int/peh-emf/publications/facts/fs304/en/}, last accessed
  on 28th November 2019.



\bibitem{chiaraviglio5g}
L.~Chiaraviglio, M.~Fiore, and E.~Rossi, {\em 5G Technology: Which Risks From
  the Health Perspective?}, pp.~37--48.
\newblock CNIT, 1~ed., 12 2019.
\newblock in Marco Ajmone Marsan, Nicola Blefari Melazzi, Stefano Buzzi, Sergio
  Palazzo, The 5G Italy Book 2019: a Multiperspective View of 5G.

\bibitem{geneve}
{\em A Gen{\`e}ve, le gel de nouvelles antennes 5G est confirm{\'e} par le
  Conseil d'Etat (in French)}.
\newblock Available at \url{https://tinyurl.com/qt9vqr2}, last accessed on 28th
  November 2019.

\bibitem{oughton2019open}
E.~J. Oughton, \textit{et al.} , ``An
  open-source techno-economic assessment framework for 5g deployment,'' {\em
  IEEE Access}, vol.~7, pp.~155930--155940, 2019.

\bibitem{matalatala2019multi}
M.~Matalatala, \textit{et al.} , ``Multi-objective optimization of
  massive mimo 5g wireless networks towards power consumption, uplink and
  downlink exposure,'' {\em Applied Sciences}, vol.~9, no.~22, p.~4974, 2019.

\bibitem{sar}
S.~Kuehn, \textit{et al.} ``Modelling of Total Exposure in Hypothetical 5G Mobile Networks for Varied Topologies and User Scenarios,'' \textit{Final Report of Project CRR-816}, Available on line at: \url{https://tinyurl.com/r6z2gqn}, June 2019 (last acc. on 1st Dec. 2019).


\bibitem{rappaport2017overview}
T.~S. Rappaport, \textit{et al.}  ``Overview of millimeter wave communications for fifth-generation
  (5g) wireless networks with a focus on propagation models,'' {\em IEEE
  Transactions on Antennas and Propagation}, vol.~65, no.~12, pp.~6213--6230,
  2017.
  
\bibitem{timeaveragedpower}
B.~Thors, \textit{et al.} ``Time-averaged realistic maximum power levels for the assessment of radio frequency exposure for 5G radio base stations using massive MIMO,'' \textit{IEEE Access}, vol.~ 5, pp.~19711--19719, 2017.

\bibitem{stone1991some}
R.~E. Stone, ``Some average distance results,'' {\em Transportation Science},
  vol.~25, no.~1, pp.~83--90, 1991.

\bibitem{larson1981urban}
R.~C. Larson and A.~R. Odoni, {\em Urban operations research}, pp.~1--573.
\newblock Prentice Hall, 1~ed., 1981.
\newblock Englewood Cliffs, NJ.

\end{thebibliography}

\end{document}